# Prompt Fission Neutron Spectra of $^{233}$U(n, F), $^{235}$U(n, F), $^{239}$Pu(n, F) and $^{240}$Pu(n,F)

Maslov V.M.
Minsk, Byelorussia

The prompt fission neutron spectra (PFNS) of fuel and fertile nuclides are of key importance for the safety/efficiency of nuclear reactors, deployed or of next generation either. The newly measured PFNS and revisited PFNS data for $^{235}$U(*n, F*), $^{239}$Pu(*n, F*) and $^{240}$Pu(*n, F*) reactions [1] strongly discard the actual PFNS evaluated data provided in available evaluated neutron data libraries [2]. The reasons for that are rather diverse. The newly measured PFNS are of double time-of-flight type. Rather wide range of the incident neutron energies, for which the outgoing prompt fission neutrons are lumped, complicates a lot the PFNS measured data fits via physical modelling, especially after the onset of the (*n,xnf*) reactions. In most cases the measured PFNS data at discrete incident neutron energies $E_{th} \lesssim E_n \lesssim 20$ MeV, as well as those measured with double time-of-flight technique were considered uncorrelated and their consistent (correlated) analyses were almost never tried upon with rare exceptions (see [3–5] and references therein).

Prompt fission neutron spectra of $^{235}$U(*n, F*), $^{239}$Pu(*n, F*) and $^{240}$Pu(*n, F*) reactions for the energy range $E_{th} \lesssim E_n \lesssim 20$ MeV might be used to predict the influence of N-odd and N-even, Z-even, *A* target nuclide and $A+1-x$ residual nuclides fissility on the emission of *x* pre-fission and post-fission neutrons [3–5]. Pre-fission neutrons exert strong influence on the observed prompt fission neutron spectra and partitioning of the fission reaction energy between total kinetic energies of fission fragments and products and their excitation energies. Initial PFNS model parameters [6] are defined by fitting $^{235}$U($n_{th}$, *f*) and $^{239}$Pu($n_{th}$, *f*) PFNS data [2].

In fact, the $(n, xnf)^{1,\ldots,x}$ and $(n, xn)^{1,\ldots,x}$ reaction competition and fissilities of the N-odd and N-even composite and cooled down, due to the emission of *x* pre-fission neutrons, fissioning residual nuclides strongly influence the observed PFNS shape and average energies of PFNS.

Exclusive neutron spectra of the $^{235}$U($n$, $xnf$)$^{1,..x}$ and $^{235}$U($n$, $xn$)$^{1,..x}$ reactions are calculated in [3−5] simultaneously with relevant neutron-induced reactions data $^{235}$U($n$, $F$), $^{235}$U($n$, $2n$), $^{234}$U($n$, $F$), $^{233}$U($n$, $F$), $^{232}$U($n$, $F$), and neutron emission spectra $^{235}$U($n$, $nX$) at $E_n$ ~14 MeV. The data on neutron emission spectra $^{238}$U($n$, $nX$) [2] were used to develop the modelling of neutron double differential distributions up to $E_n$ ~18 MeV. The same approach we pursued for $^{239}$Pu+$n$ and $^{240}$Pu+$n$ interactions.

Exclusive pre-fission neutron spectra of $^{239}$Pu($n$, $xnf$)$^{1,\ldots,x}$ , $^{240}$Pu($n$, $xnf$)$^{1,\ldots,x}$ reactions are calculated simultaneously with relevant neutron-induced reactions cross sections $^{240}$Pu($n$, $F$), $^{239}$Pu($n$, $F$), $^{239}$Pu($n$, $2n$), $^{238}$Pu($n$, $F$), $^{237}$Pu($n$, $F$) and $^{236}$Pu($n$, $F$) [3−5]. Neutron emission spectra $^{239}$Pu($n$, $nX$) at $E_n$ ~14 MeV and inelastic neutron scattering cross sections are reproduced as well.

Initial PFNS model parameters for $^{240}$Pu($n$, $F$) PFNS are fixed via description of the spontaneous fission neutron spectra (SFNS) of $^{240}$Pu($sf$). The $^{240}$Pu ($n$, $xnf$)$^{1,..x}$ pre–fission neutron spectra, total kinetic energy of fission fragments, average prompt fission neutron number and observed PFNS of $^{240}$Pu($n$, $F$) are predicted [3−5].

Measured PFNS data are measured using as a reference spontaneous fission neutron spectra (SFNS) of $^{252}$Cf($sf$). Recent measurements [7] supported previous estimates of $^{252}$Cf($sf$) SFNS which were used in [2], however PFNS measurements [2] still may be susceptible to the systematic biases of various origin. These systematic errors are cancelled in measured ratios of PFNS. Available measured ratios of PFNS $^{239}$Pu($n$, $F$)/$^{235}$U($n$, $F$), $^{239}$Pu($n$, $F$)/$^{238}$U($n$, $F$) and $^{235}$U($n$, $F$)/$^{238}$U($n$, $F$) for $E_{\rm th} < E_n <$ 20 MeV [8], strongly support calculated PFNS of $^{235}$U($n$, $F$), $^{239}$Pu($n$, $F$), $^{238}$U($n$, $F$) [3−5].

The PFNS of $^{235}$U($n$, $F$), $^{239}$Pu($n$, $F$), $^{238}$U($n$, $F$) [3−5] are mostly strongly different when compared with PFNS of available evaluated neutron data libraries [2], though the same data base [2] might be used.

## PFNS of $^{235}$U($n$, $F$)

Our estimate of PFNS is based on all then available measured data [2]. PFNS of $^{235}$U($n$,$F$) in ENDF/B-VIII.1 [1] at thermal neutron energy $E_n$ ~$E_{th}$ (Fig. 1) simulates our estimate presented at Nuclear Data conference of 2010 at JeJu (Korea) [6].

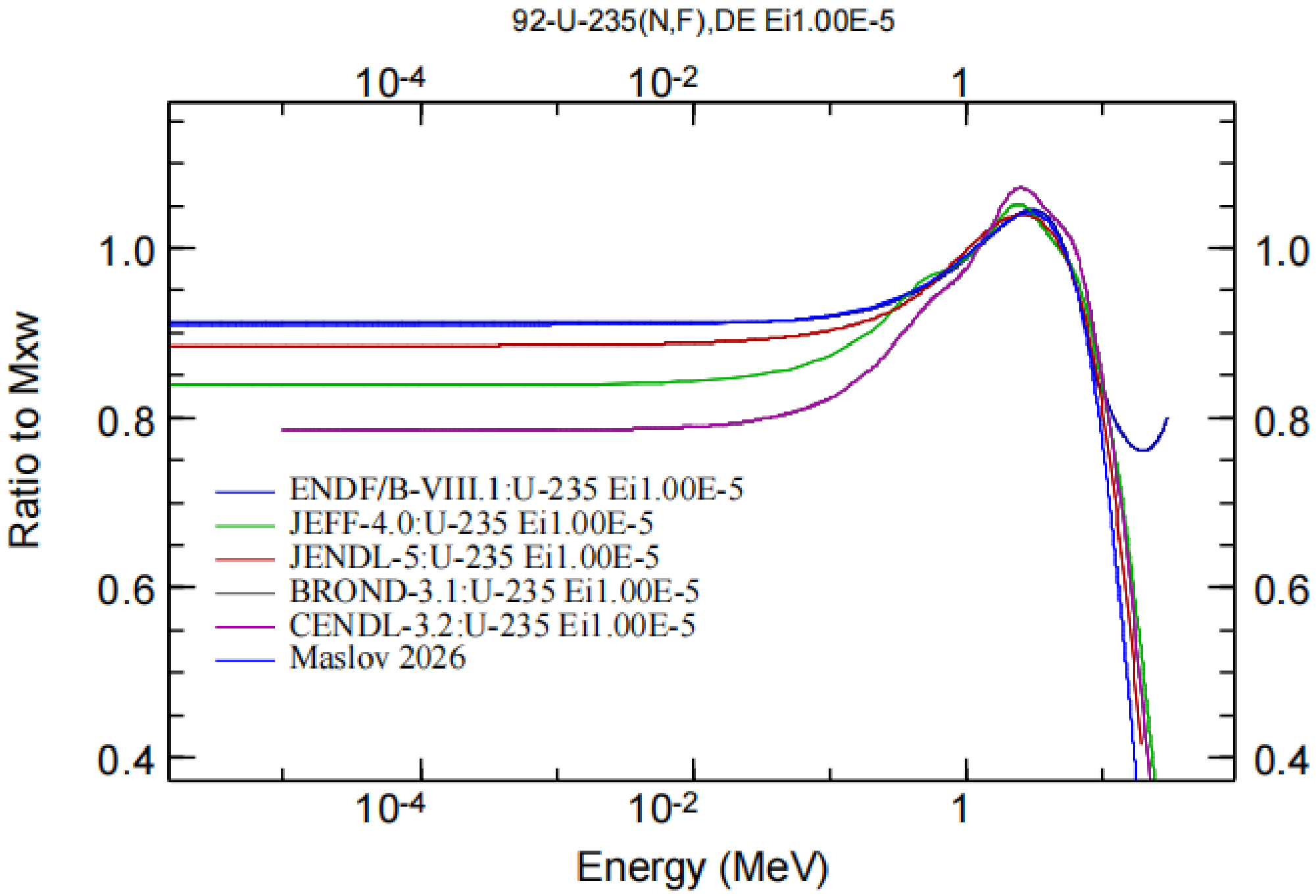

Fig.1 PFNS of $^{235}$U($n$,$F$) at thermal point $E_n$ ~$E_{th}$

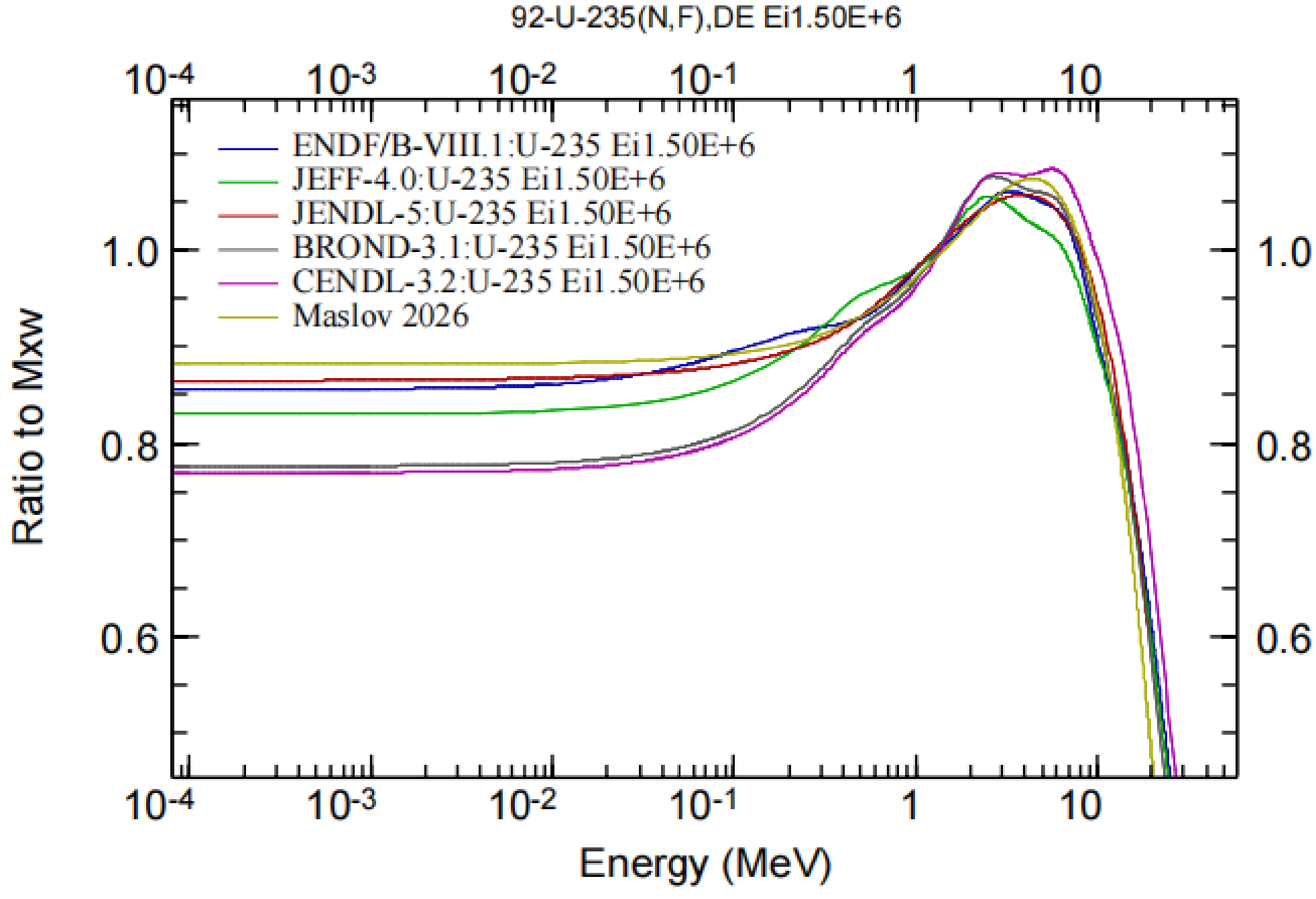

Fig. 2 PFNS of $^{235}$U($n$,$F$) at $E_n$ ~1.5 MeV

At $E_n$ ~1.5 MeV only ENDF/B-VIII.1 evaluation follows newly measured data [1] and is compatible with our estimate [5] (see Fig. 2). The variation of present PFNS at $E_n$ ~1.5 MeV and $E_n$ ~ $E_{th}$ MeV is due to the increase of the incident neutron energy.

At higher $E_n$, after the onset of $^{235}$U($n$,$nf$) reaction, the evaluations, which are brand-new and those which are of historical interest, are drastically discrepant with the measured data [1] and each other (see Figs. 3-5). As explained in [3–5] the contribution of $^{235}$U($n$,$nf$) reaction to the observed PFNS much depends on the competition with the $^{235}$U($n$,$2n$) reaction. The largest relative contribution to the observed $^{235}$U($n$,$F$) PFNS happens at $E_n$ ~6.5 MeV [3, 5]. That peculiarity is missing in available evaluated libraries [2], but is in new measurements [1]. The cooling down of the fission fragments due to pre-fission neutron emission is pronounced as a variation of PFNS “steepness” at $\varepsilon > E_{nnf1}$, the cut-off energy of the $^{235}$U($n$,$nf$)[1] pre-fission neutron. As explained in [3−5] we provide the partial contributions of $^{235}$U(*n*,*nf*) and $^{235}$U(*n*,*2nf*) reactions to the observed PFNS, which comes from the exclusive pre-fission neutron spectra and relevant fission fragments. At 15 MeV the reasonable shape of pre-fission neutrons $^{235}$U(*n*,*nf*) and $^{235}$U(*n*,*2nf*) is evident in our calculations only. It confirmed by the newly measured PFNS [2].

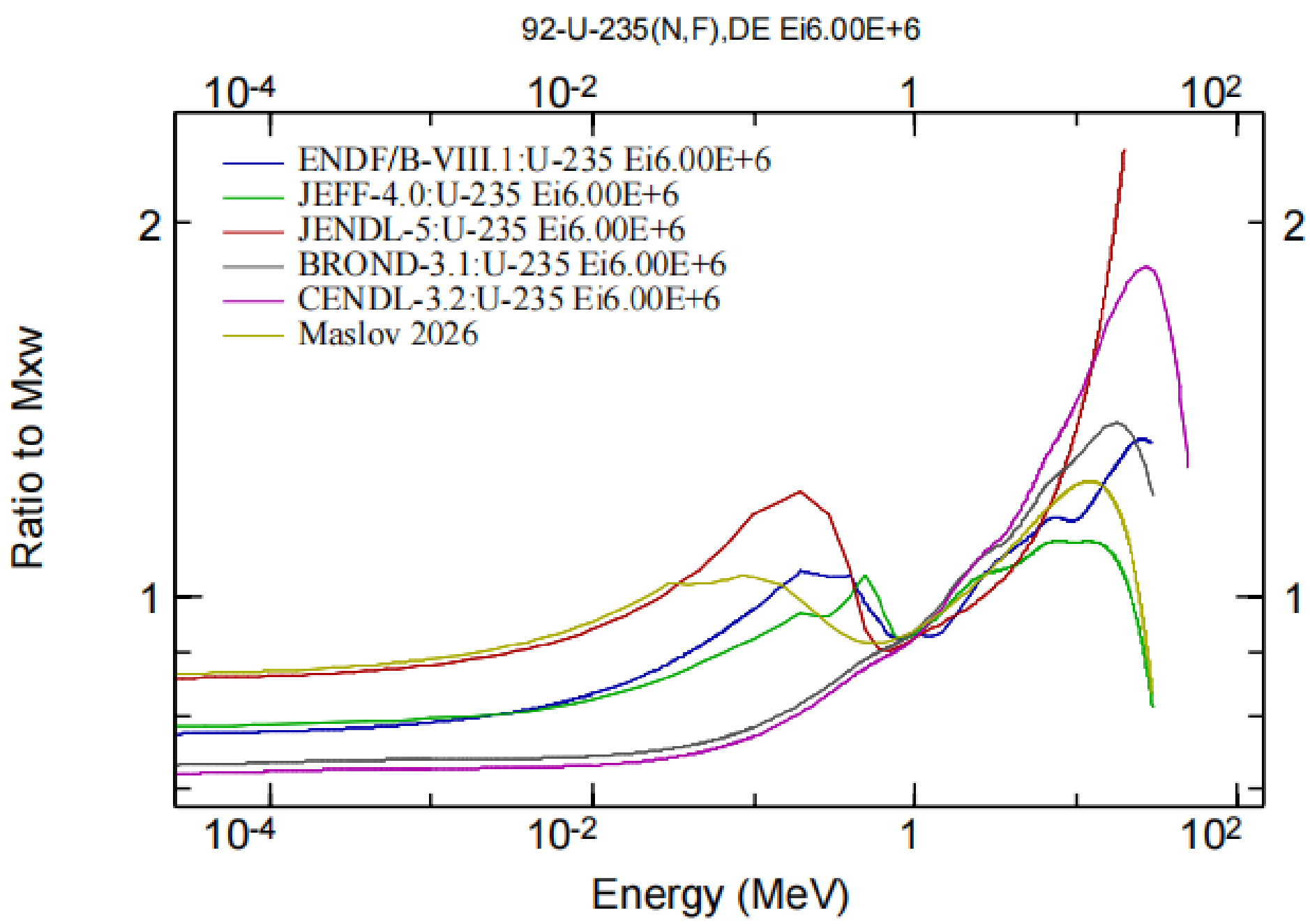

Fig. 3 PFNS of $^{235}$U(*n*,*F*) at $E_n$ ~6.0 MeV

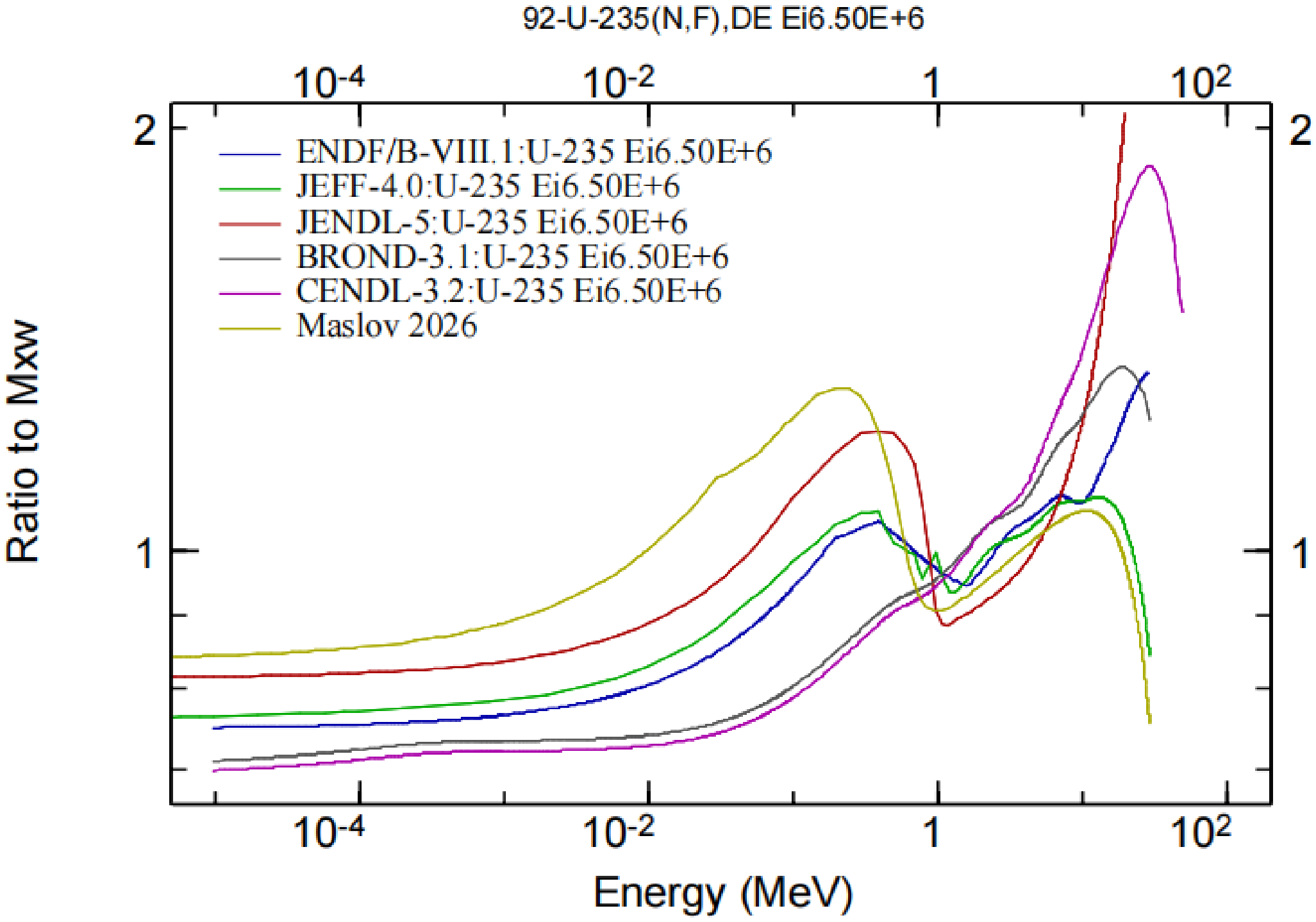

Fig. 4 PFNS of $^{235}$U($n$,$F$) at $E_n$ ~6.5 MeV

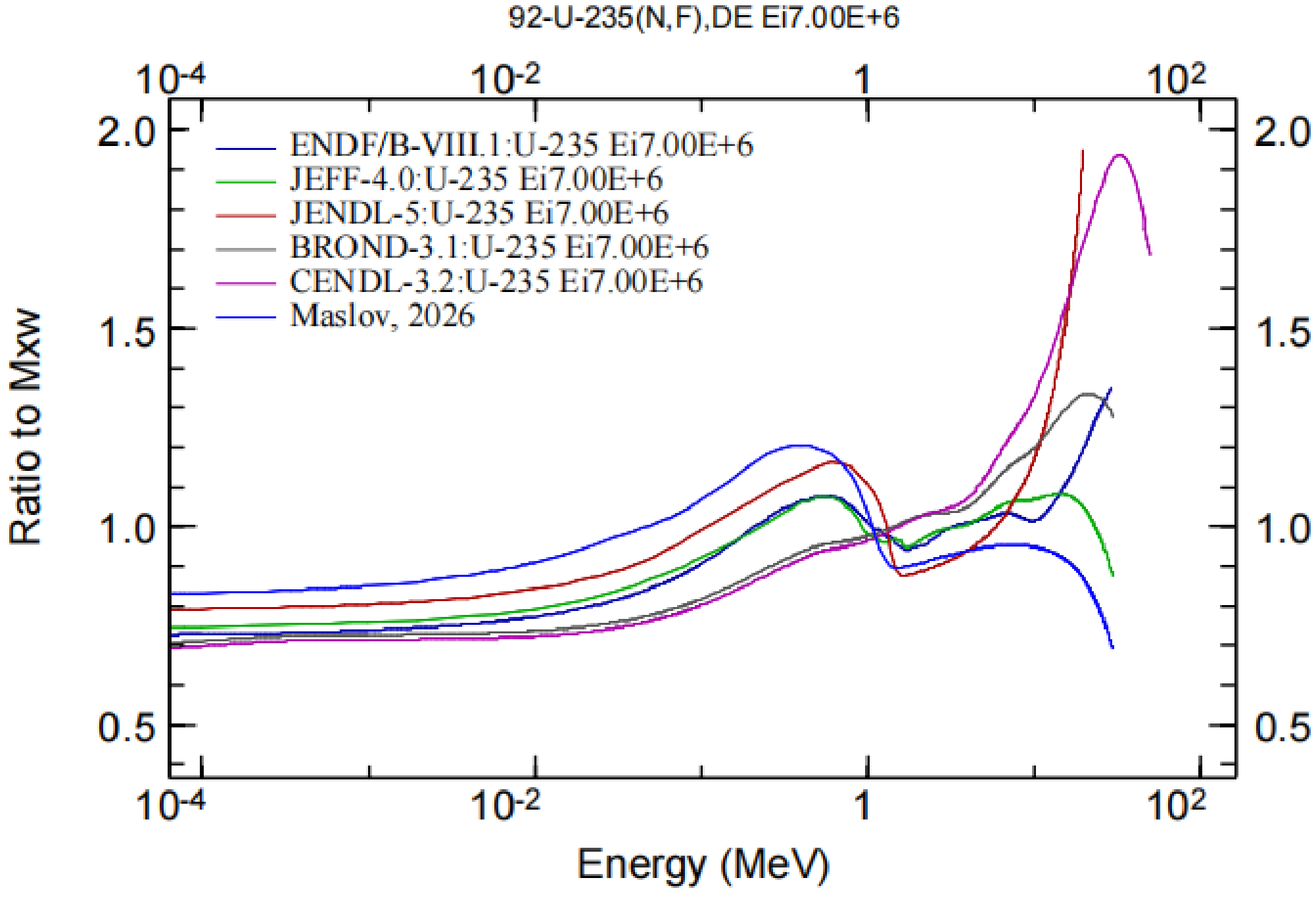

Fig. 5. PFNS of $^{235}$U($n$,$F$) at $E_n$ ~7.0 MeV

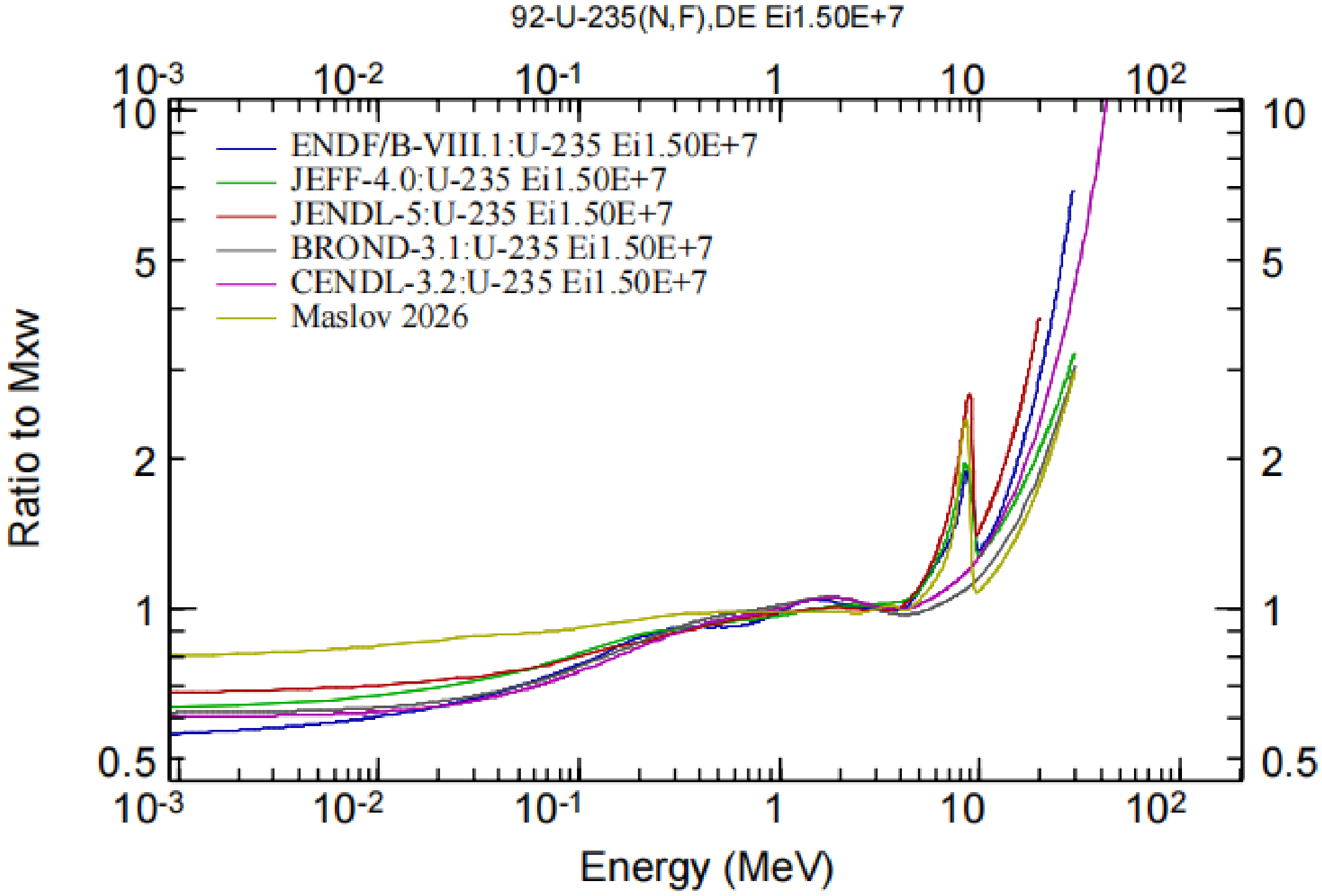

Fig. 6 PFNS of $^{235}$U($n$,$F$) at $E_n$ ~15.0 MeV

## PFNS of $^{239}$Pu($n$, $F$)

PFNS of $^{239}$Pu($n$,$F$) in ENDF/B-VIII.1 at thermal point $E_n \sim E_{th}$ (Fig. 7) roughly simulates our estimate presented at Nuclear Data conference (2010) at JeJu (Korea) [6].

Our estimate of PFNS of $^{239}$Pu($n$,$F$) at $E_n$ ~1.5 MeV [3, 4] (see Fig. 8) is compatible with available measured ratios of PFNS $^{239}$Pu($n$, $F$)/$^{235}$U($n$, $F$), $^{239}$Pu($n$, $F$)/$^{238}$U($n$, $F$)) [8].

Contribution of exclusive pre-fission neutron spectra of $^{239}$Pu($n$, $xnf$)$^{1,\ldots,x}$ , to the observed PFNS of $^{239}$Pu($n$, $F$)is much lower than in case of $^{235}$U($n$, $F$) reaction. The largest relative contribution to the observed $^{239}$Pu($n$,$F$) PFNS happens at $E_n$ ~6 MeV [3, 5]. Exclusive neutron spectra of $^{239}$Pu($n$, $xnf$)$^{1,\ldots,x}$ reactions are calculated simultaneously with relevant neutron-induced reactions cross sections $^{239}$Pu($n$, $F$), $^{239}$Pu($n$, $2n$), $^{238}$Pu($n$, $F$), $^{237}$Pu($n$, $F$) and $^{236}$Pu($n$, $F$) [3−5].

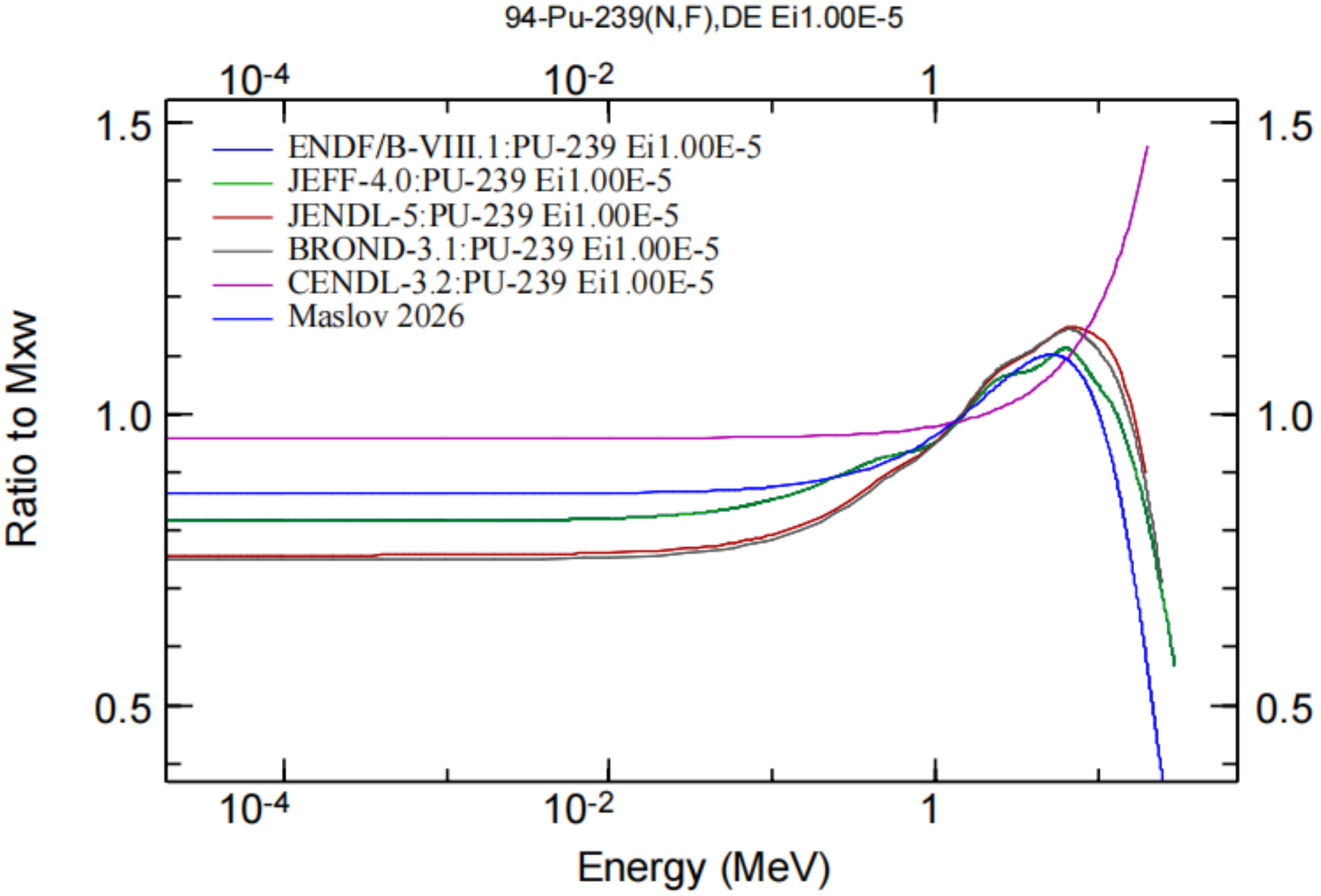

Fig.7 PFNS of $^{239}$Pu($n,f$) at thermal point $E_n$ ~$E_{th}$

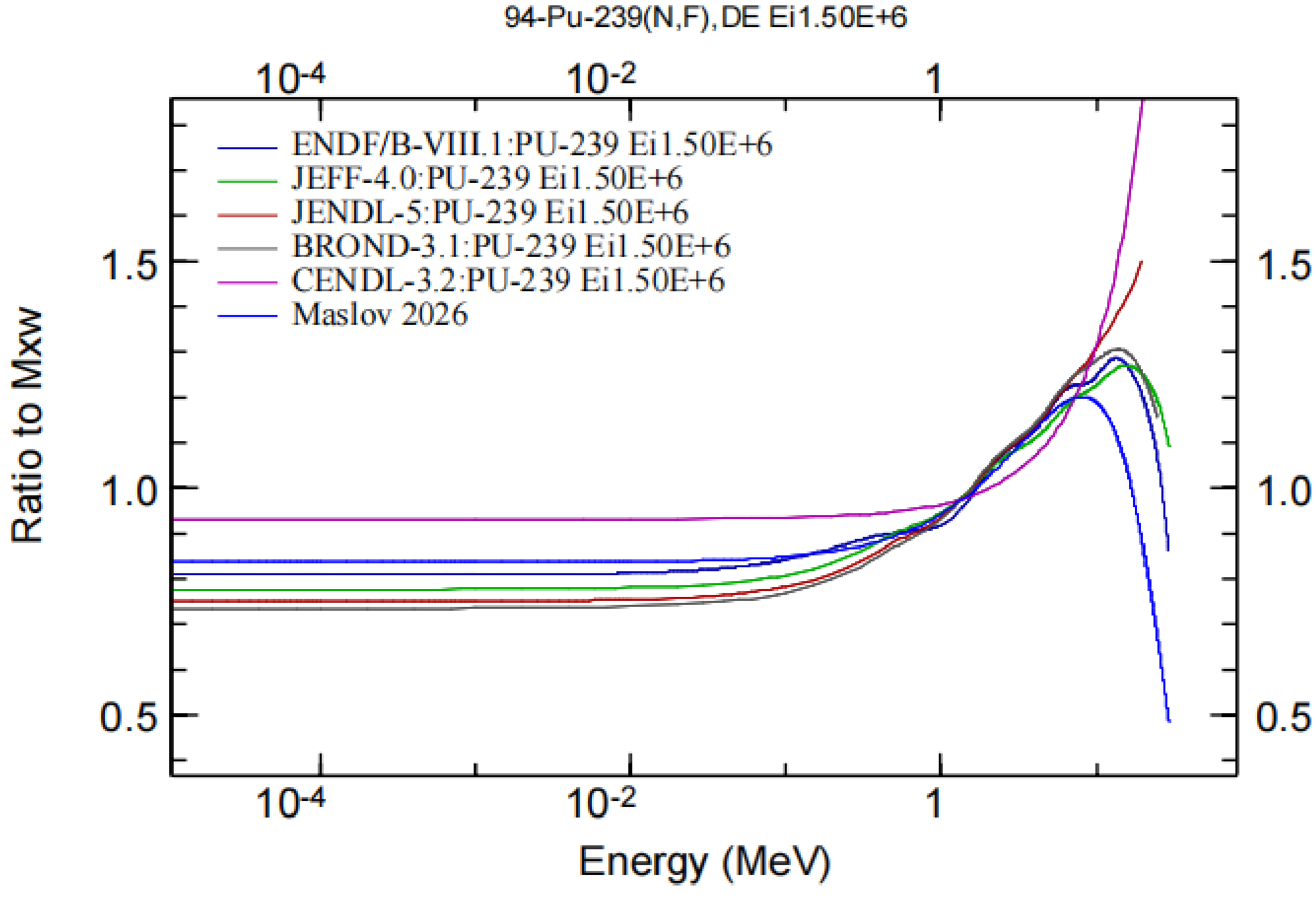

Fig.8 PFNS of $^{239}$Pu($n,f$) $E_n$ ~1.5 MeV

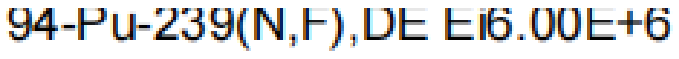

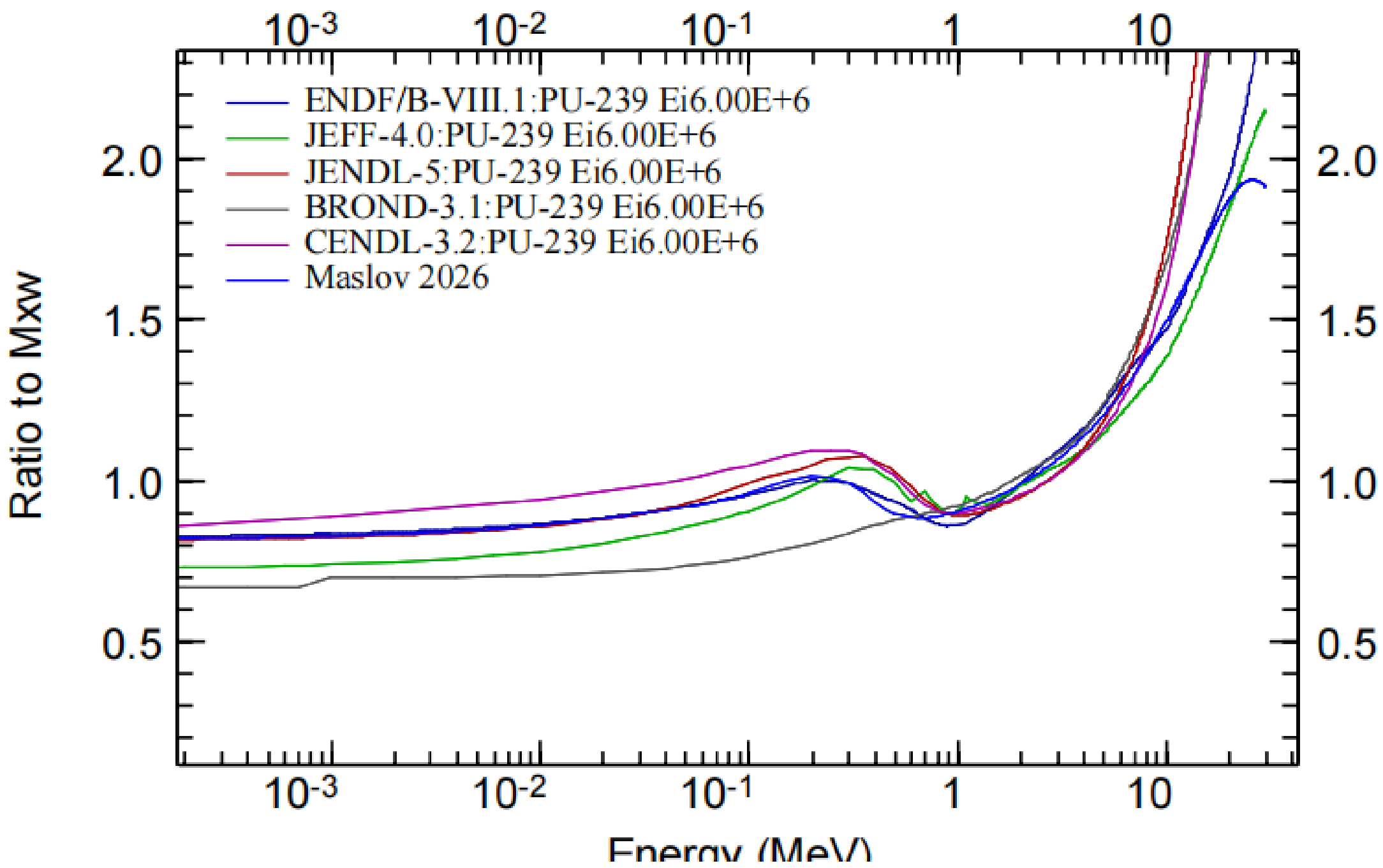

Fig. 9. PFNS of $^{239}$Pu($n$,$F$) at $E_n$ ~6.0 MeV

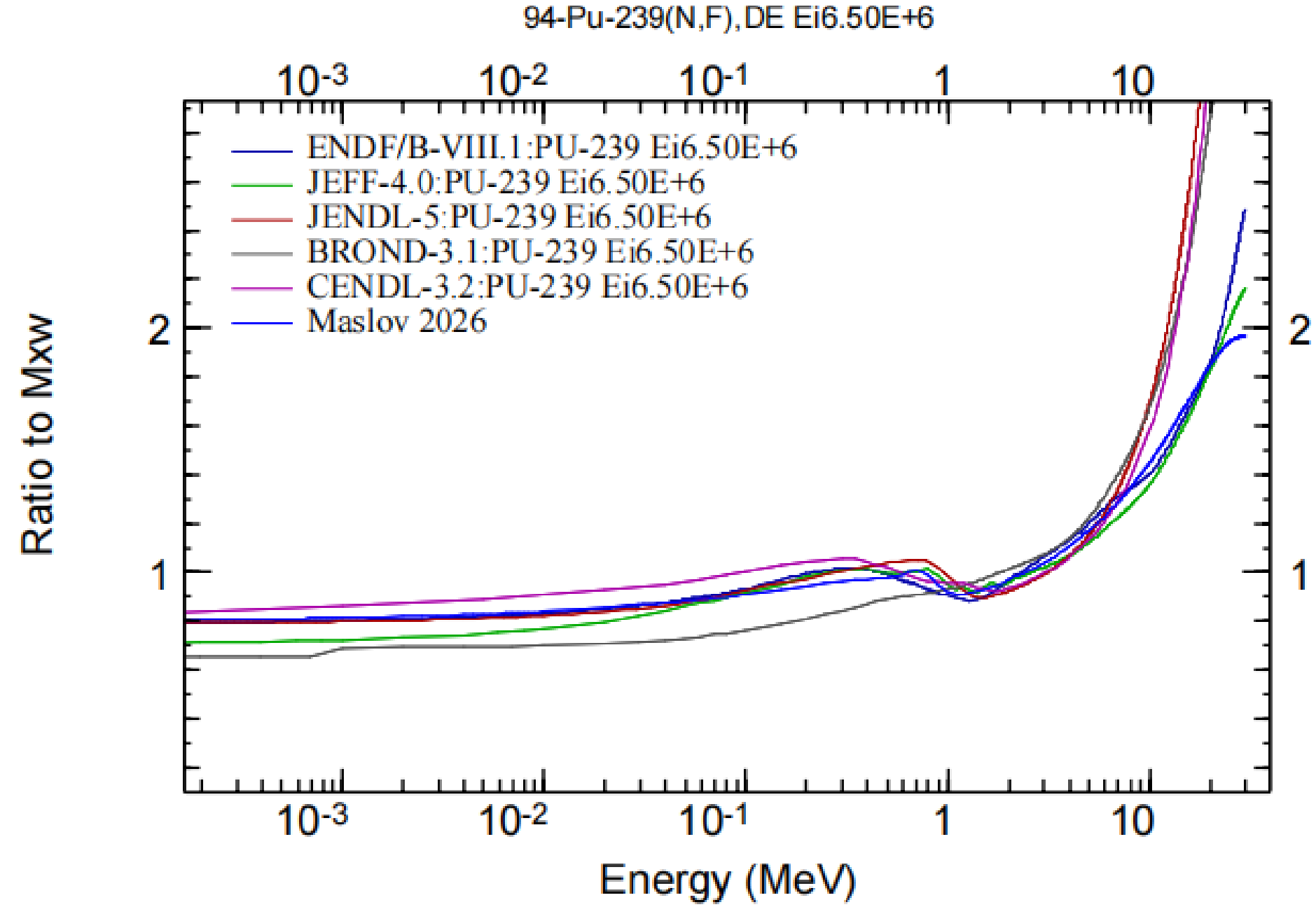

Fig. 10. PFNS of $^{239}$Pu($n$,$F$) at $E_n$ ~6.5 MeV

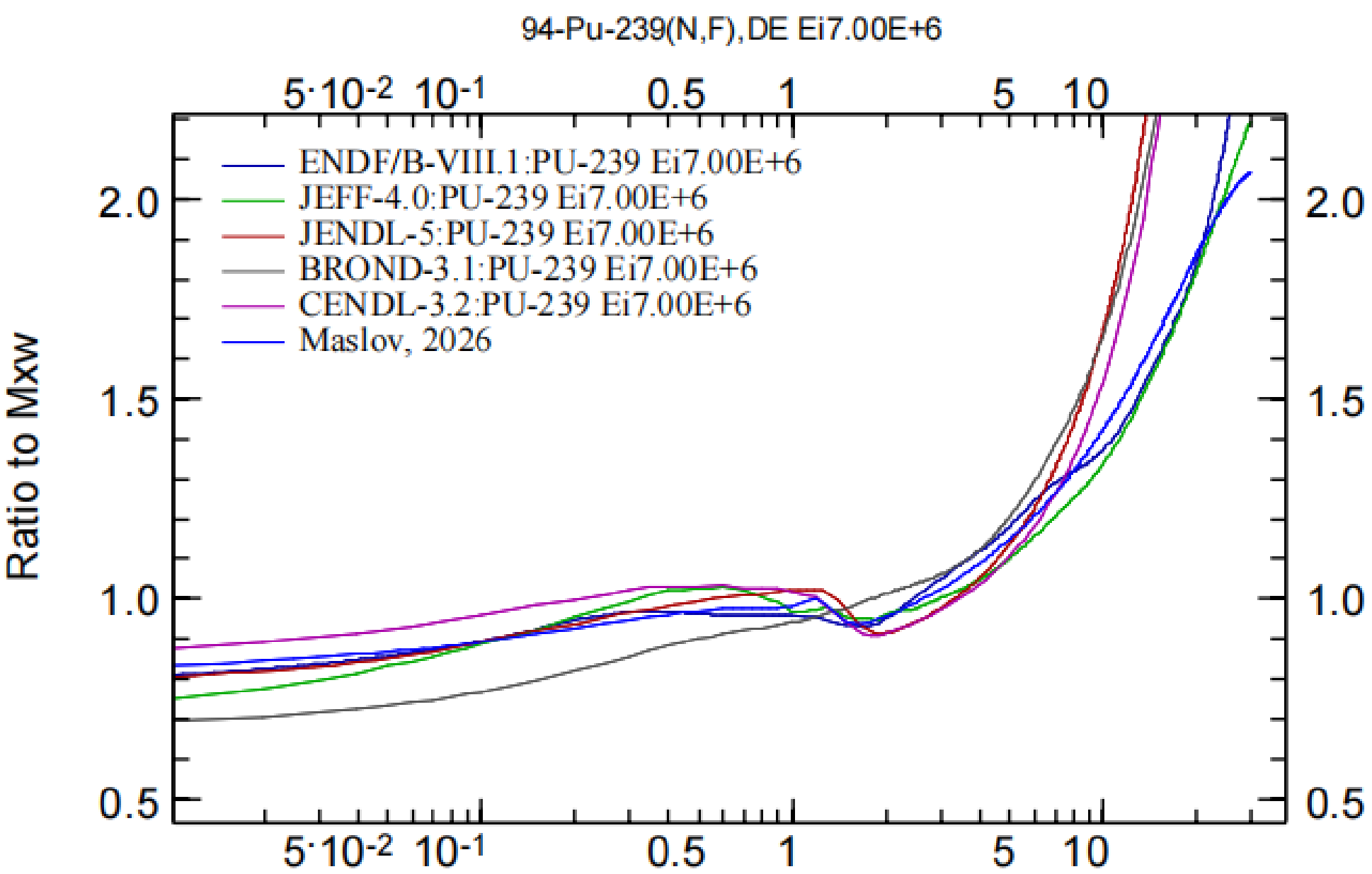

Fig. 11. PFNS of $^{239}$Pu($n$,$F$) at $E_n$ ~7.0 MeV

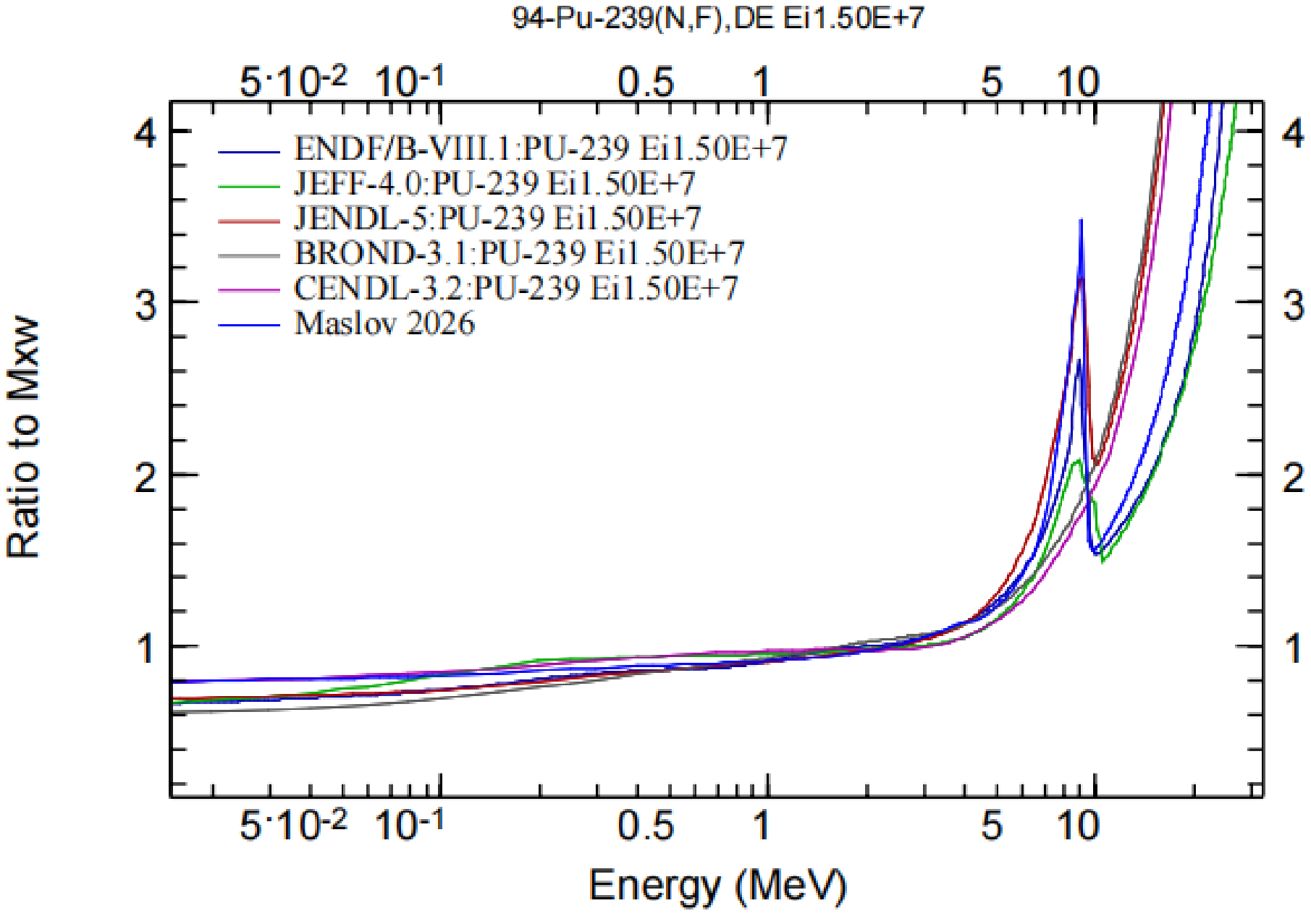

Fig. 12. PFNS of $^{239}$Pu($n$,$F$) at $E_n$ ~15 MeV

Contribution of exclusive pre-fission neutron spectra of $^{239}$Pu($n$, $xnf$)$^{1,\ldots,x}$ to the observed PFNS of $^{239}$Pu($n$, $F$)is much lower than in case of $^{235}$U($n$, $F$) reaction. The largest relative contribution to the observed $^{239}$Pu($n$,$F$) PFNS happens at $E_n$ ~6 MeV [3, 5]. Exclusive neutron spectra of $^{239}$Pu($n$, $xnf$)$^{1,\ldots,x}$ reactions are calculated simultaneously with relevant neutron-induced reactions cross sections $^{239}$Pu($n$, $F$), $^{239}$Pu($n$, $2n$), $^{238}$Pu($n$, $F$), $^{237}$Pu($n$, $F$) and $^{236}$Pu($n$, $F$) [3−5].

## PFNS of $^{240}$Pu($n$, $F$)

Predicted PFNS of $^{240}$Pu($n$, $F$) [5] at $E_n \lesssim 20$ MeV are supported by measurements at neutron source LANSCE [2].

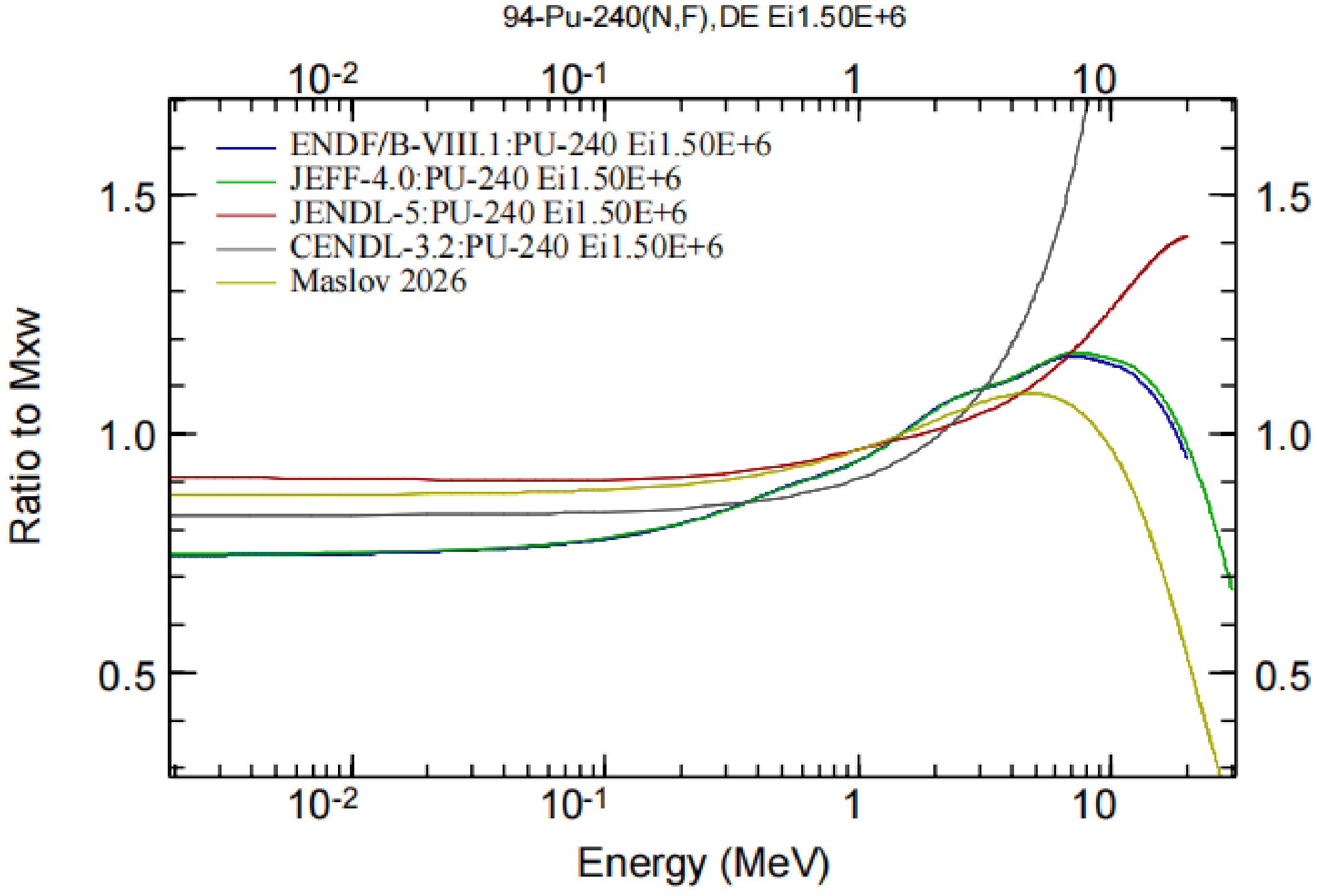

Fig. 13. PFNS of $^{240}$Pu($n$,$F$) at $E_n$ ~1.5 MeV

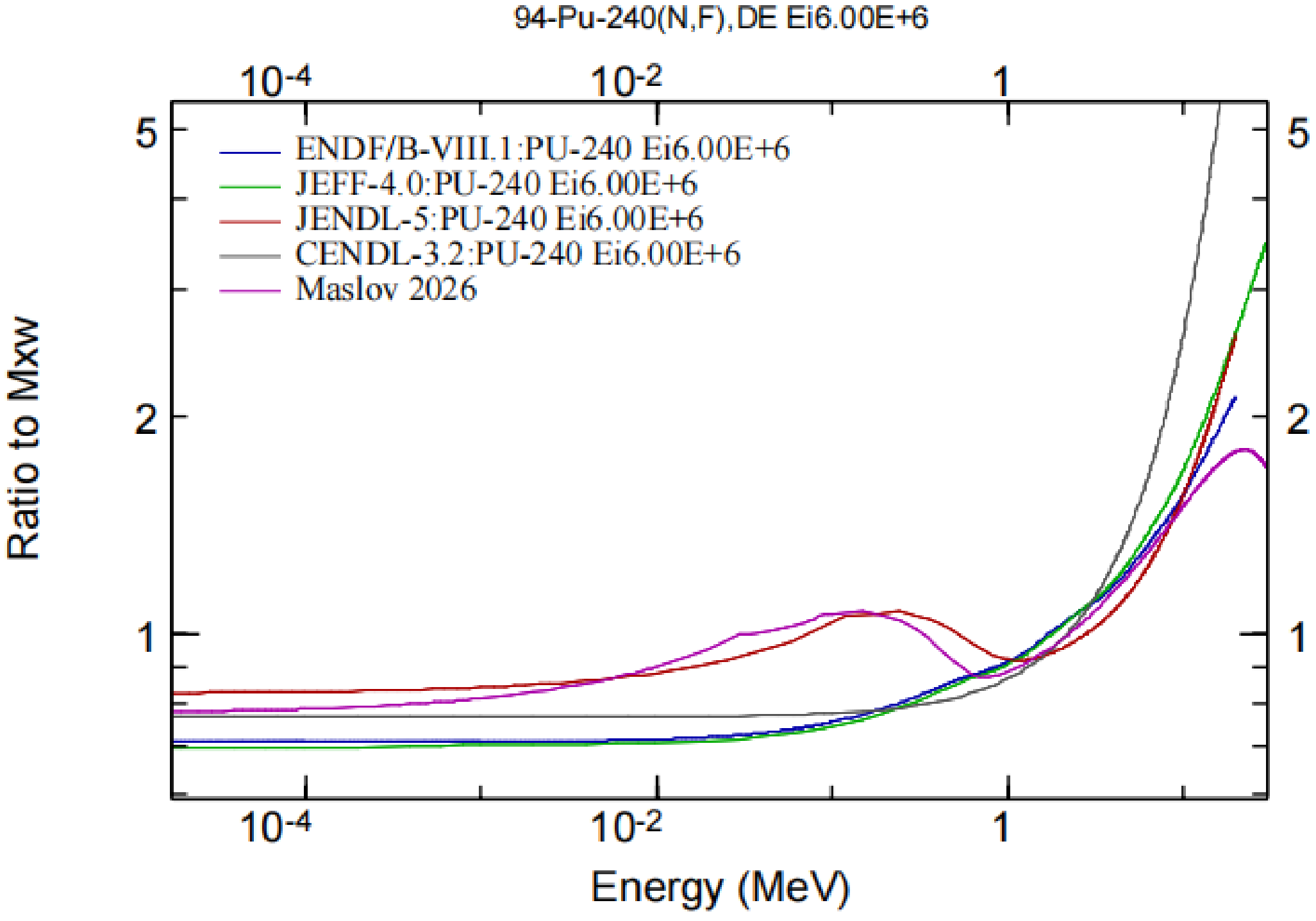

Fig. 14. PFNS of $^{240}$Pu($n$,$F$) at $E_n$ ~6.0 MeV

Exclusive neutron spectra of $^{240}$Pu($n$, $xnf$)$^{1,\ldots}$are calculated simultaneously with relevant neutron-induced reactions cross sections $^{239}$Pu($n$, $F$), $^{239}$Pu($n$, $2n$), $^{238}$Pu($n$, $F$), and $^{237}$Pu($n$, $F$) and [3−5]. Contribution of exclusive pre-fission neutron spectra of $^{240}$Pu($n$, $xnf$)$^{1,\ldots,x}$ , to the observed PFNS of $^{240}$Pu($n$, $F$) is much lower than in case of $^{235}$U($n$, $F$) reaction, but higher than in case of $^{239}$Pu($n$, $F$) reaction. The largest relative contribution to the observed $^{240}$Pu($n$,$F$) PFNS happens at $E_n$ ~6 MeV [3, 5] (see Figs. 14, 15, 16). For N-even $^{240}$Pu target nuclides the contribution of exclusive neutron spectra of $^{240}$Pu($n$, $xnf$)$^{1,\ldots,x}$ reactions should be compared with those for $^{238}$U and $^{232}$Th [4]. In latter cases the relevant contributions are much larger [4].

After the onset of $^{240}$Pu($n$,$nf$) reaction the evaluations, which are brand-new and those which are of historical interest only, are drastically discrepant with the measured data [2] and each other (see Figs. 14, 15, 16, 17, 18, 19). As explained in [3−5] the contribution of $^{240}$Pu($n$,$nf$) reaction to the observed PFNS much depends

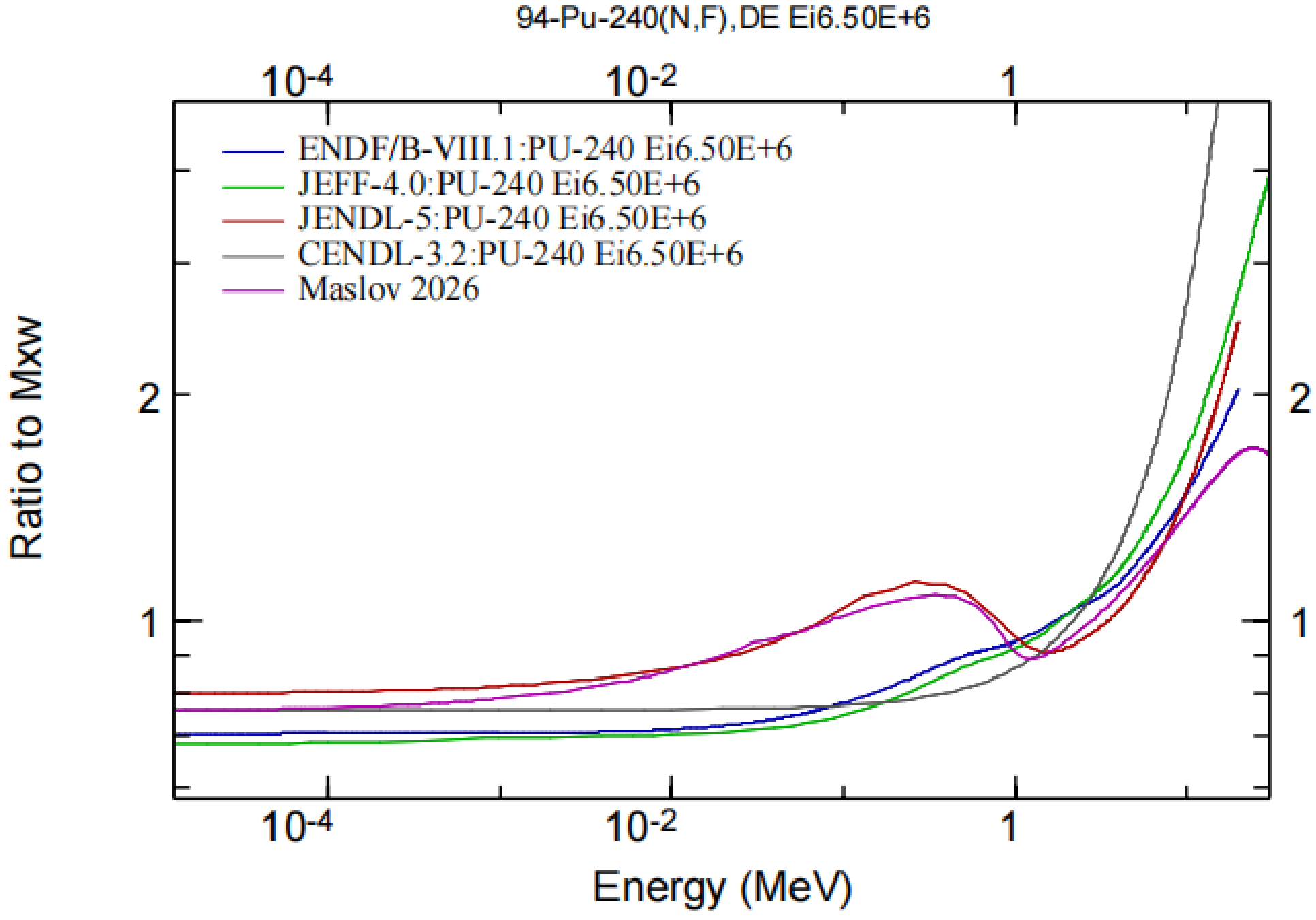

Fig. 15. PFNS of $^{240}$Pu($n$,$F$) at $E_n$ ~6.5 MeV

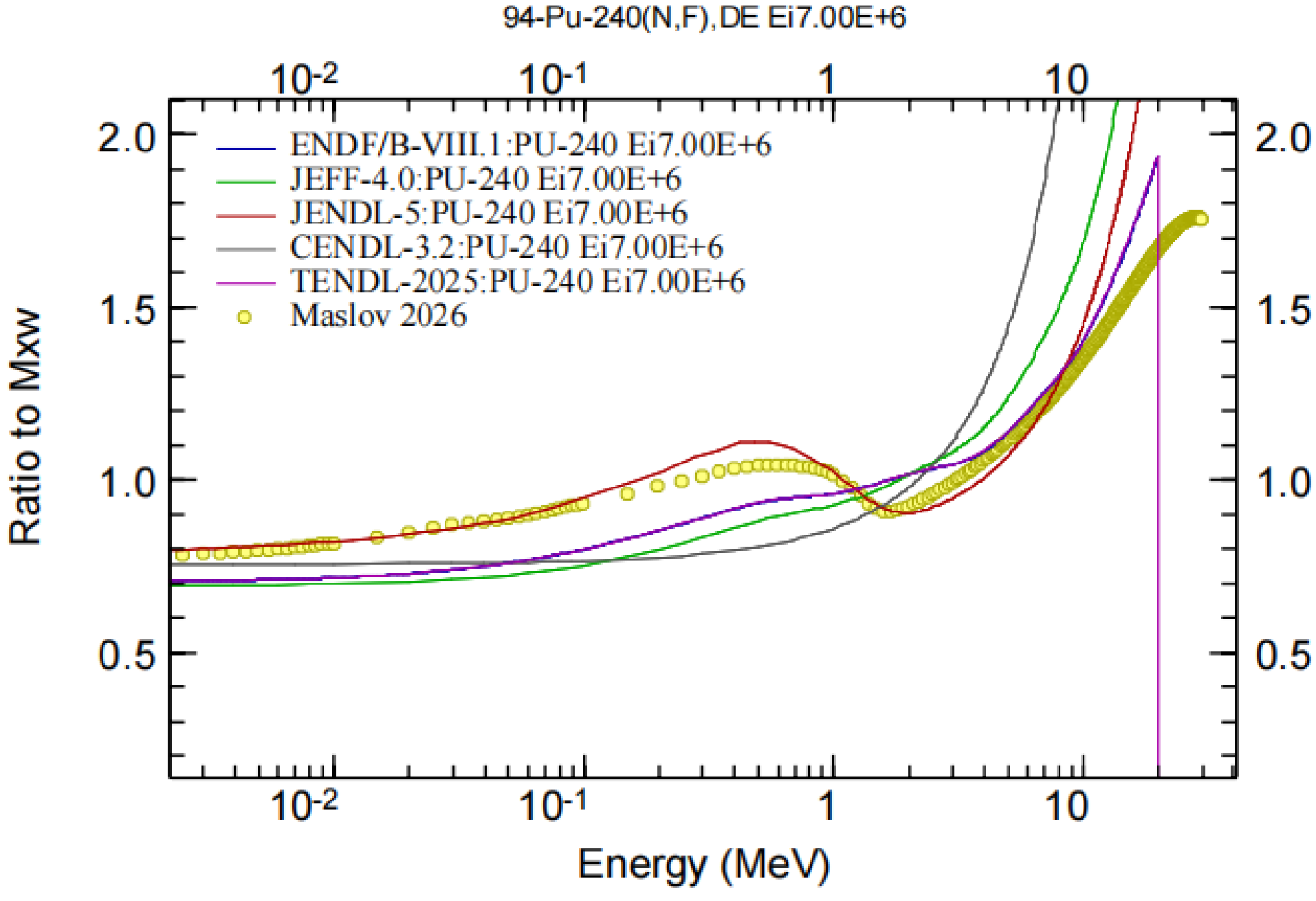

Fig. 16. PFNS of $^{240}$Pu($n$,$F$) at $E_n$ ~7 MeV

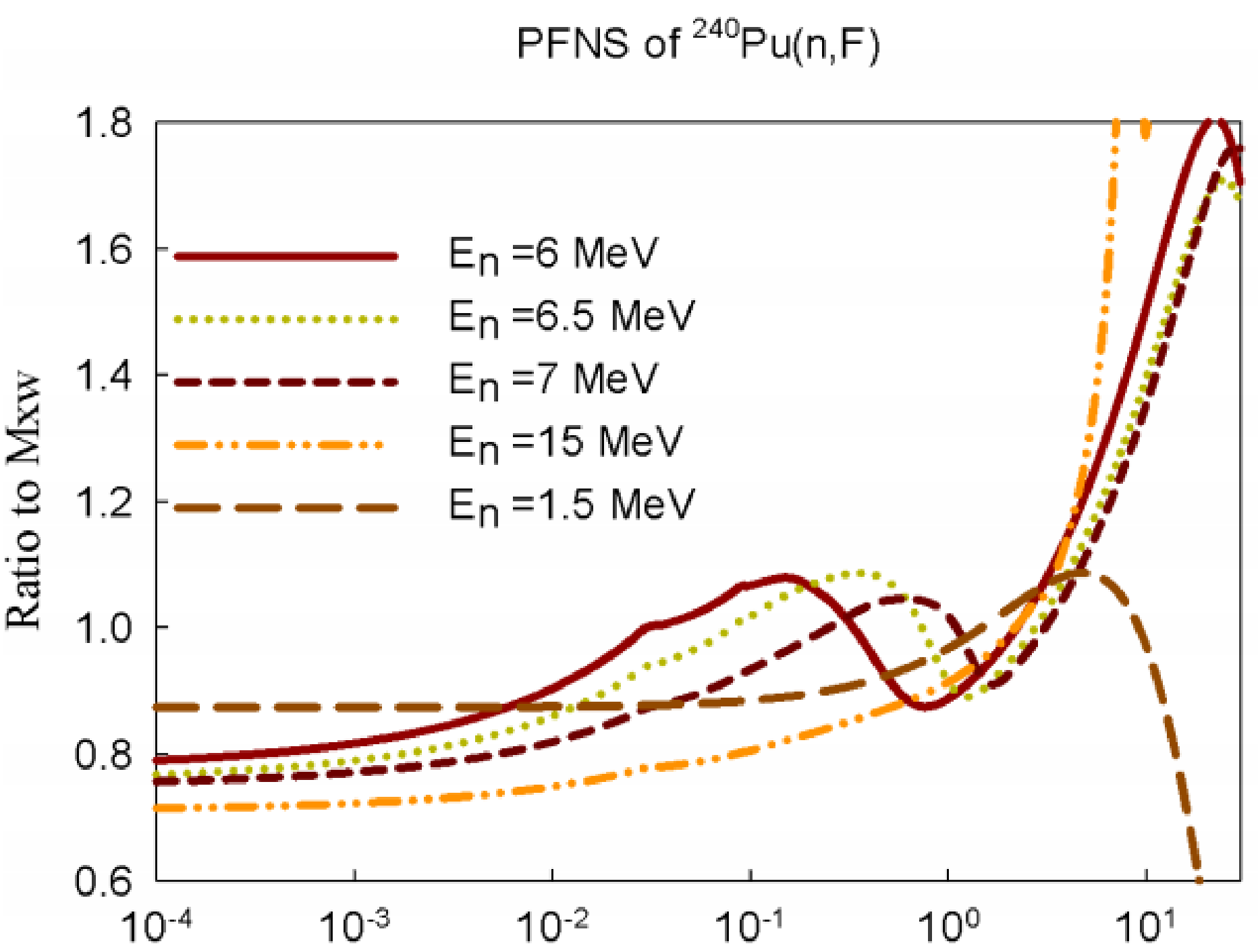

Fig. 17. PFNS of $^{240}$Pu($n$,$F$)

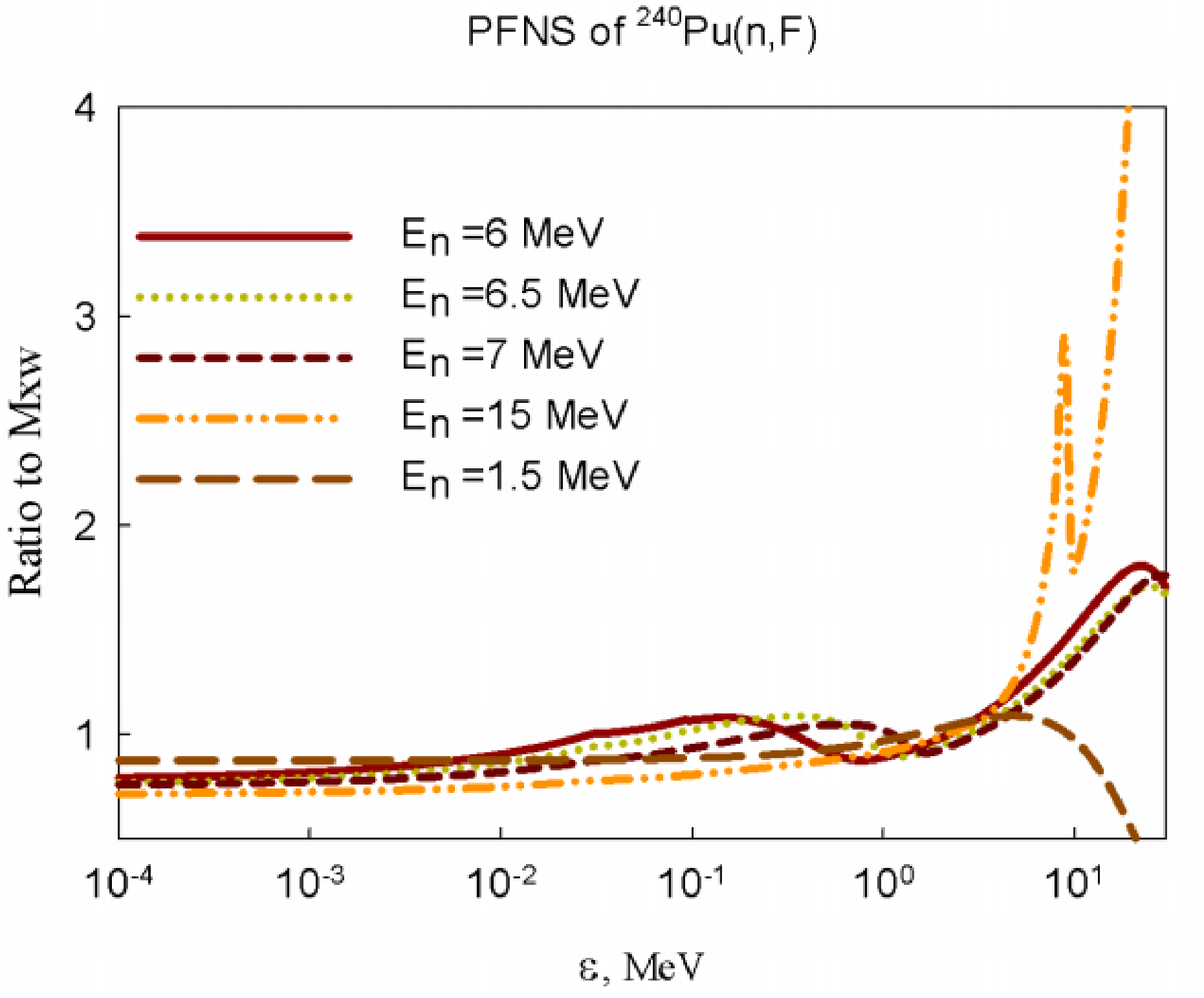

Fig. 18. PFNS of $^{240}$Pu($n$,$F$)

on the competition with $^{240}$Pu($n$,$2n$) reaction and energies of pre-fission neutrons.

Figures 17, 18 demonstrates strong influence of pre-fission neutrons and excitation energies on the PFNS shape. The variation of the PFNS shape in the vicinity of the $^{240}$Pu($n$,$nf$) reaction threshold provokes the variation of the average energy of PFNS of $\langle E\rangle$~150 keV. At $E_n$ ~15 MeV the pre-fission neutron of $^{240}$Pu$(n,nf)^1$ reaction has a soft low energy tail and hard high energy tail with a cutoff energy of ~10.5 MeV [3, 4].

Analysis of prompt fission neutron spectra of $^{235}$U($n$, $F$), $^{240}$Pu($n$, $F$) and $^{240}$Pu($n$, $F$) evidenced correlations of a number of observed fission data structures with $(n,xnf)^{1\ldots x}$ pre-fission neutrons. Pre-fission neutron spectra turned out to be quite soft as compared with neutrons emitted by excited fission fragments. The net effect of these peculiarities is the drop of $\langle E\rangle$ in the vicinity of the thresholds of ($n$, $xnf$) reactions. The amplitude of the variation of $\langle E\rangle$ is much higher in case of $^{240}$Pu($n$, $F$) as compared with that of $^{239}$Pu($n$, $F$). These peculiarities are due to varying emissive fission ($n$,$xnf$) contributions in $^{239}$Pu($n$, $F$) and $^{240}$Pu ($n$, $F$) reactions.

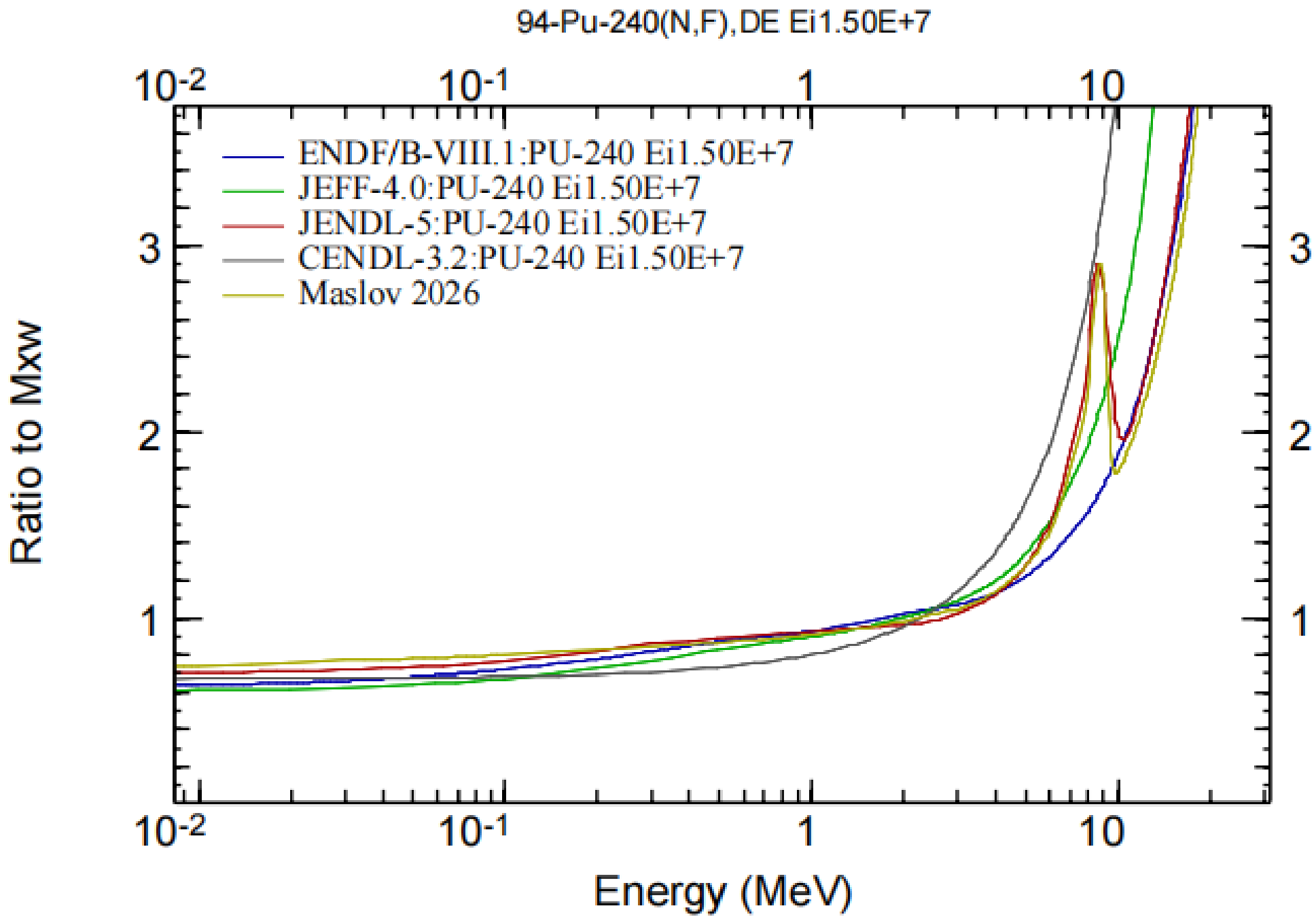

Fig. 19 PFNS of $^{240}$Pu($n$,$F$) at $E_n$ ~15 MeV

The structures in average total kinetic energies of fission fragment TKE $E_F^{pre}$ and fission products $E_F^{post}$ are shown to be correlated with the emission of (*n, xnf*) neutrons due to cooling down of the respective fissioning nuclides [3, 5].

The model for the PFNS calculation [3-5] was tuned for some twenty years. Analysis of prompt fission neutron spectra of $^{239}$Pu(*n, F*) started in [9, 10], it revealed/interpreted correlations of a number of observed data structures with $^{239}$Pu(*n,xnf*)$^{1...x}$ pre-fission neutrons [11]. Pre-fission neutron spectra turned out to be quite soft as compared with neutrons emitted by excited fission fragments. The net effect is the decrease of $\langle E \rangle$ in the vicinity of the thresholds of $^{238,239,240,241,242}$Pu(*n, xnf*) reactions [12]. The amplitude of the variation of $\langle E \rangle$ in case of $^{240,242}$Pu(*n, F*) and $^{241}$Pu(*n, F*) is higher as compared with that of $^{239}$Pu(*n, F*). The correlation of PFNS shape with angles of emission of (*n, xnf*)$^1$ neutrons and emissive fission contributions for $^{238,239,240,241,242}$Pu(*n, F*) reaction cross section is established. The angular anisotropy of exclusive pre-fission neutron (*n, xnf*)$^1$ spectra strongly influences the PFNS shapes and their average energies $\langle E \rangle$. These peculiarities are due to varying emissive fission (*n,xnf*) contributions in $^{239}$Pu(*n, F*) and $^{238,240,241,242}$Pu(*n, F*) reactions. Calculated ratio of $\langle E \rangle$ for “forward” and “backward” emission of pre-fission neutrons steeply rises with the increase of average energies of exclusive pre-fission neutron spectra $^{238,239,240,241,242}$Pu (*n,xnf*)$^{1...x}$.

The calculated anisotropy of pre-fission neutrons of $^{240,241,242}$Pu(*n,xnf*) reaction is a bit higher than in case of $^{239}$Pu(*n, F*) reaction. That might be due to correlation of anisotropy of pre-fission neutrons, which increases the average energy of pre-fission neutron spectra, with contribution of emissive fission (*n, nf*) to the observed fission cross sections, PFNS and angular anisotropy of neutron emission spectra.

# PFNS of $^{233}$U(*n, F*)

In case of $^{233}$U($n$,$F$) and $^{235}$U($n$,$F$) tractions the PFNS have some similar and different features. Observed peculiarities in PFNS, TKE, $\nu_p(E_n)$ correlate with the emission of pre-fission ($n$,$xnf$) neutrons, as predicted for the $^{233}$U($n$,$F$) and $^{233}$U($n$,$xnf$) and earlier for $^{235}$U($n$,$F$) and $^{235}$U($n$,$xnf$) [13, 14]. The correlation of PFNS shape and emissive (($n$,$xnf$)) fission contribution to the observed fission cross section for $^{233}$U($n$,$F$) and $^{235}$U($n$,$F$) reactions is much stronger than in case of $^{238,240,241,242}$Pu targets. Amplitude of dips in $\langle E \rangle$ of $^{233}$U($n$,$F$) PFNS is quite similar to that observed in PFNS of $^{235}$U($n$,$F$) reaction, notwithstanding the appreciable differences of $^{233}$U($n$,$xnf$) and $^{235}$U($n$,$xnf$) reaction contributions to the observed fission cross sections $^{233}$U($n$,$F$) and $^{235}$U($n$,$F$), respectively. That is explained by relatively large contributions of partial $\nu_{px}(E_{nx})$ as compared with $\nu_{pre}(E_n)$, number of pre-fission neutrons, for the reaction $^{233}$U($n$,$F$). PFNS of $^{233}$U($n$,$F$) are more hard than those of $^{235}$U($n$,$F$) PFNS, but softer than those of $^{239}$Pu($n$,$F$). It might be argued that correct estimate of the exclusive pre-fission ($n$,$xnf$) neutron spectra and modelling of spectra of neutrons emitted from excited fission fragments gives a robust prediction of PFNS for $^{233}$U($n$,$F$) [15] for incident neutron energies $E_n$~ $E_{th}$–20 MeV with a precision and reliability comparable to those attained for $^{235}$U($n$,$F$)

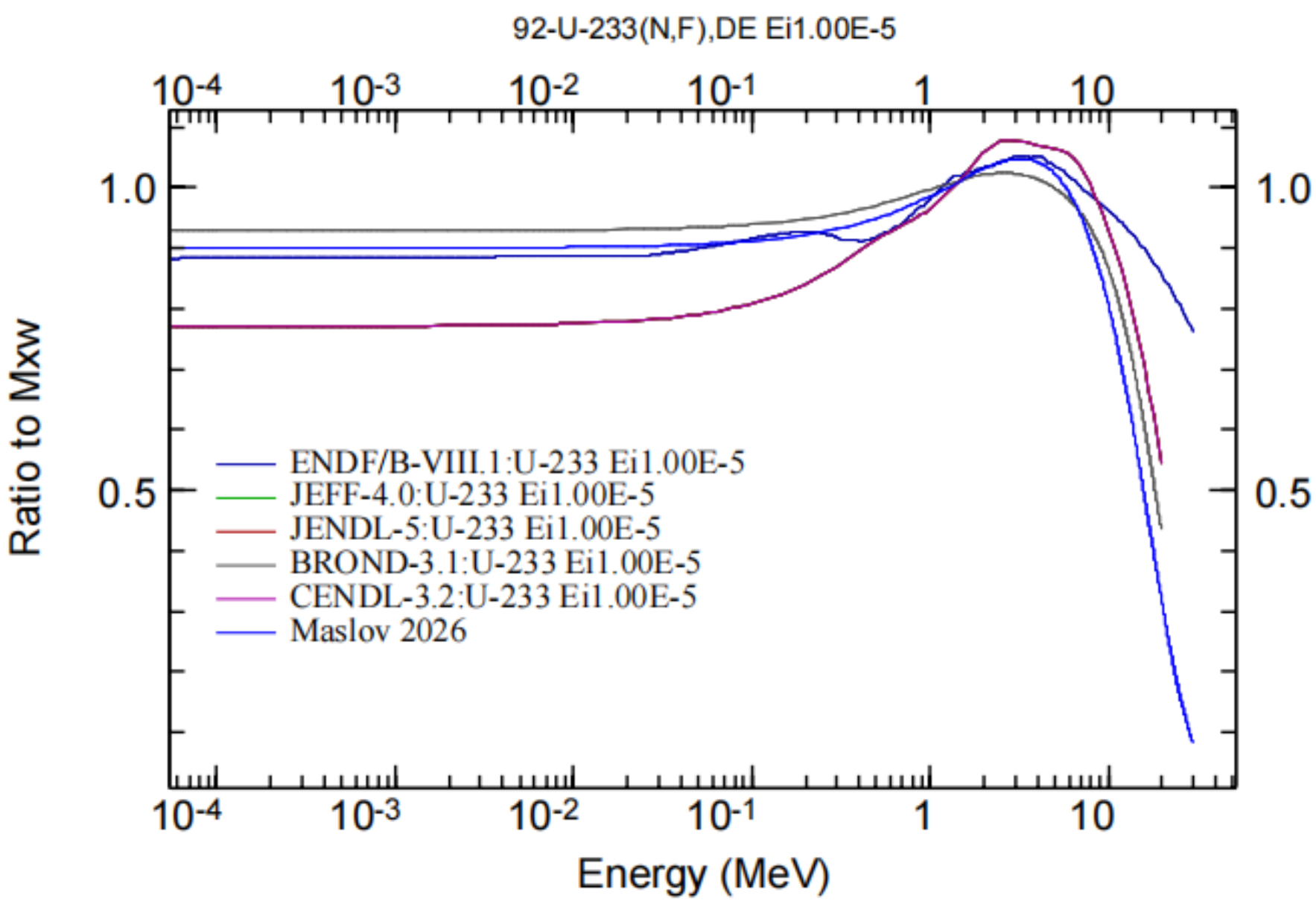

Fig.20 PFNS of $^{233}$U($n$,$f$) at thermal point $E_n$ ~$E_{th}$

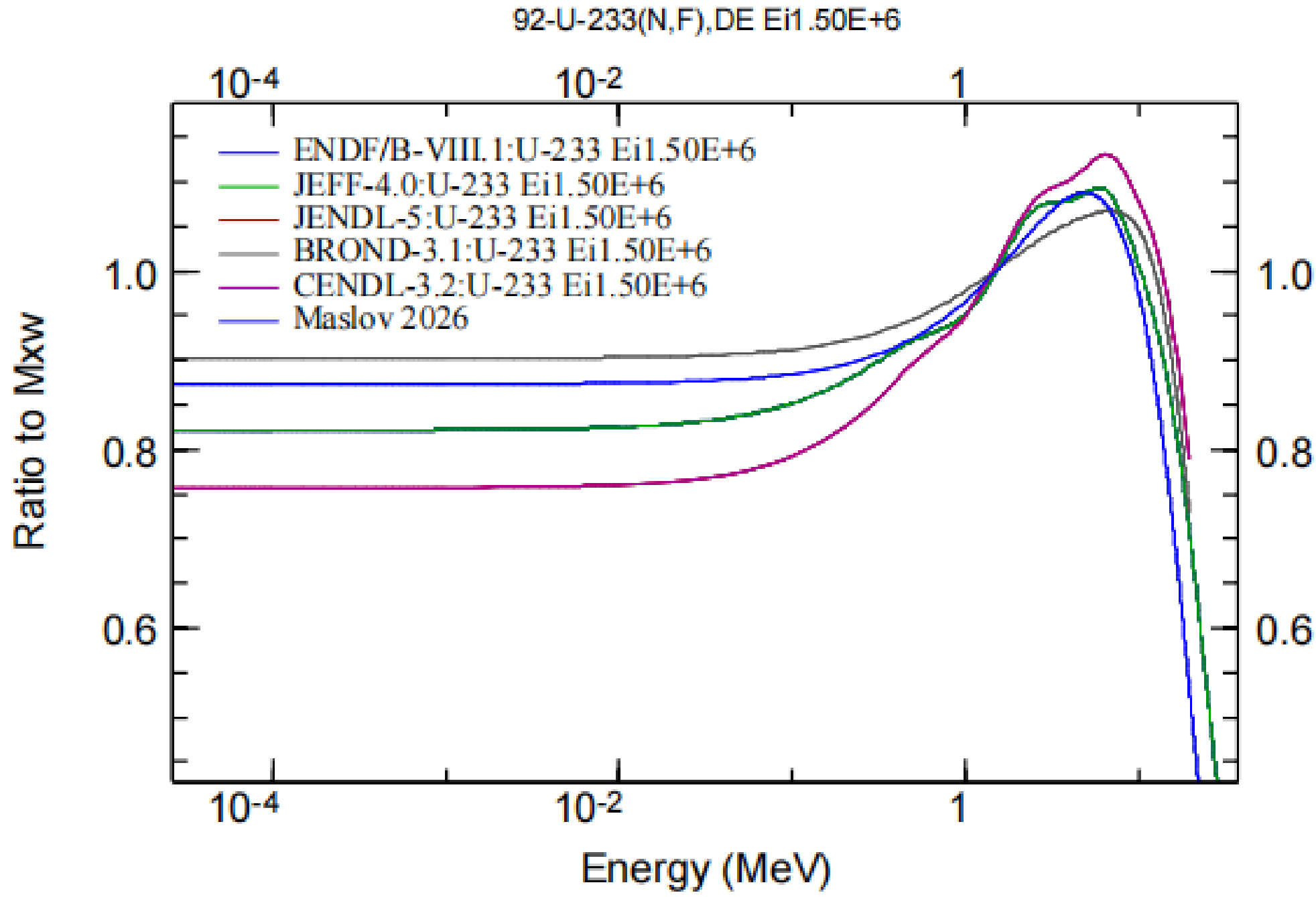

Fig.21 PFNS of $^{233}$U($n,f$) at $E_n$ ~1.5 MeV

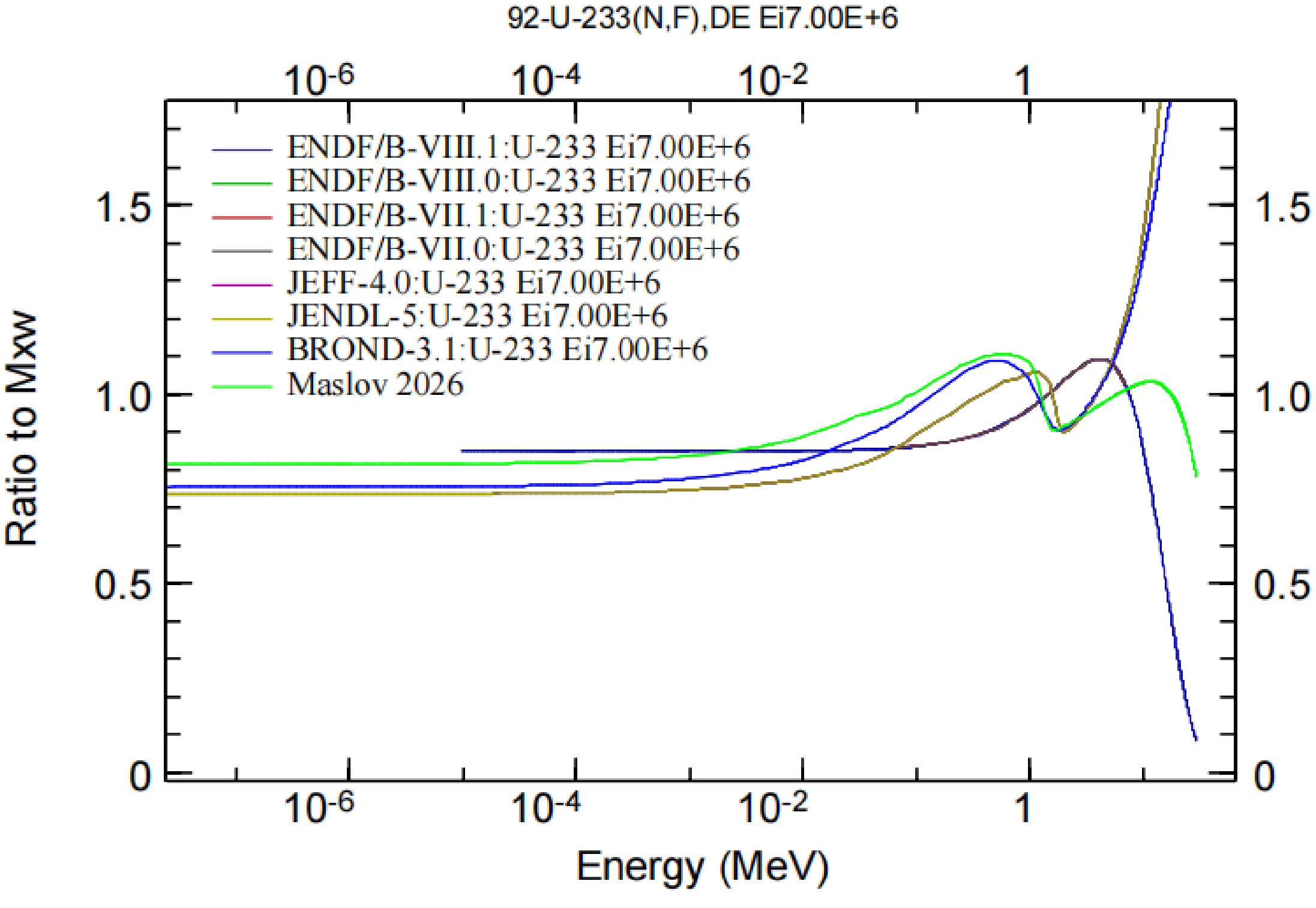

Fig.22 PFNS of $^{233}$U($n,f$) at $E_n$ ~7 MeV

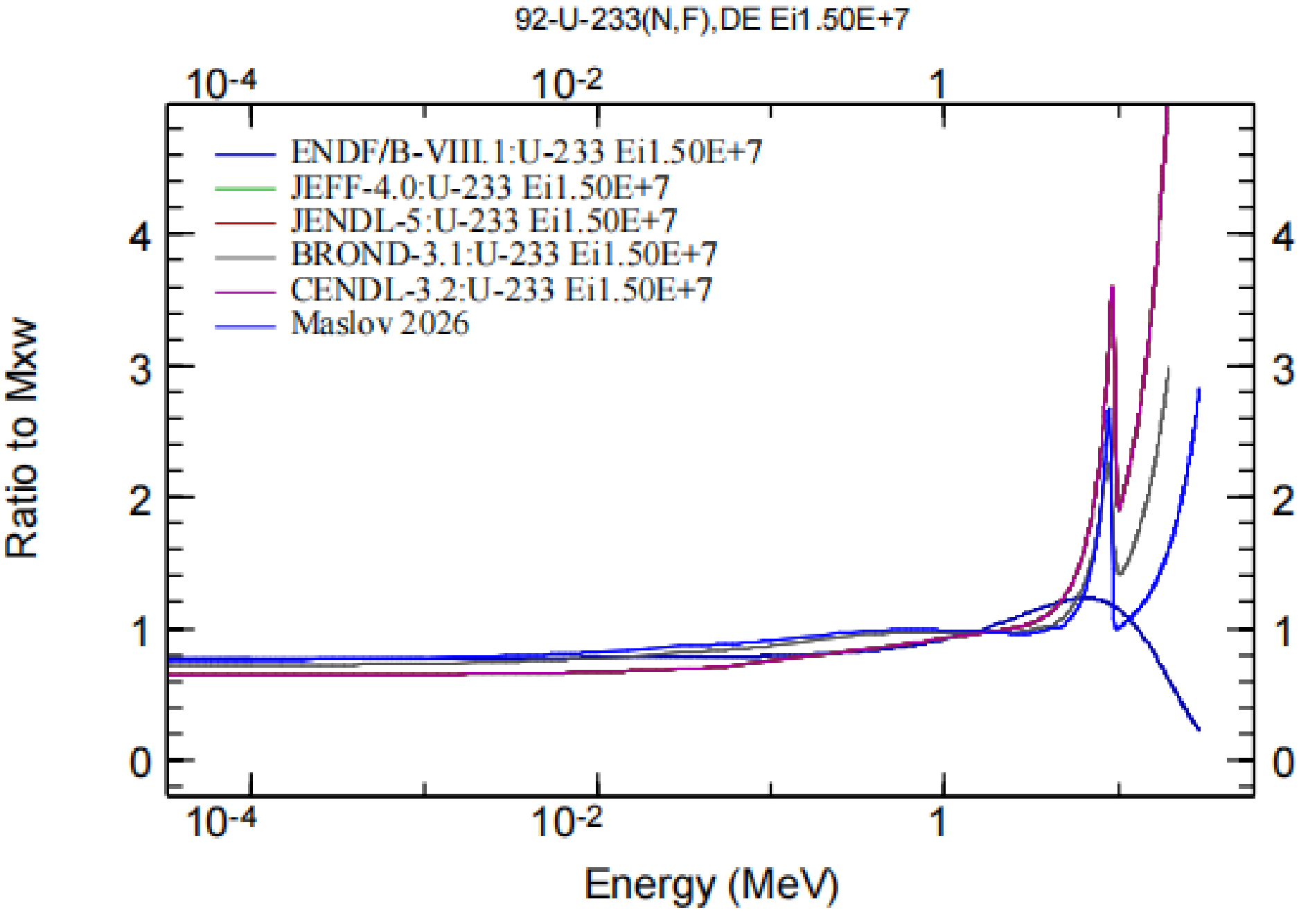

Fig.23 PFNS of $^{233}$U($n$,$f$) at $E_n$ ~15 MeV

PFNS. The anomalous asymmetry of observed prompt fission neutrons is predicted in [16].

PFNS of $^{233}$U(n,F) is produced with the models polished in case of $^{239}$Pu and $^{235}$U targets [3, 5].

PFNS of $^{233}$U($n$,$f$) at thermal point $E_n$ ~$E_{th}$ of ENDF/B-VIII.1 roughly simulates our estimate of 2010 at ND conference at JeJu. At $E_n$ ~1.5 MeV ENDF/B-VIII.1 replaced previously adopted JENDL evaluation with former ENDF/B numerical data. That is a step in right direction, as is evident with our calculations.

At $E_n$ ~7 MeV the reasonable rate of pre-fission neutrons is in our calculations only. No wonder in ENDF/B-VIII.1 they replaced previously adopted JENDL evaluation with former ENDF/B numerical data, though the latter PFNS shape is rather arbitrary.

At $E_n$ ~15 MeV the reasonable shape of pre-fission neutrons is in our calculations only. No wonder in ENDF/B-VIII.1 they replaced previously adopted JENDL evaluation with former ENDF/B numerical data, though the latter PFNS shape is again rather arbitrary. It seems preliminary 233U PFNS data of LANSCE are already available.

$^{233}$U(*n*,*F*) PFNS data of LANSCE are round the corner ($^{232}$U is a problem) [15].

## Conclusions

Actual PFNS evaluated data of available evaluated neutron data libraries [2] still are in severe disagreement with new measured PFNS and revisited PFNS data for $^{235}$U(*n, F*), $^{239}$Pu(*n, F*) and $^{240}$Pu(*n, F*) reactions [1]. Rather wide range of the incident neutron energies, for which the outgoing prompt fission neutrons are lumped need robust physical modelling, especially after the onset of the (*n*,*xnf*) reactions. Consistent analysis of measured PFNS data at discrete incident neutron energies $E_{th} \lesssim E_n \lesssim 20$ MeV, as well as those measured with double time-of-flight technique, needs further efforts (see [3−5] and references therein).